\documentstyle[12pt]{article}

\newcommand{\nsect}{\setcounter{equation}{0}
\def\theequation{\thesection.\arabic{equation}}\section}
\def\simlt{\mathrel{\lower2.5pt\vbox{\lineskip=0pt\baselineskip=0pt
           \hbox{$<$}\hbox{$\sim$}}}}
\def\simgt{\mathrel{\lower2.5pt\vbox{\lineskip=0pt\baselineskip=0pt
           \hbox{$>$}\hbox{$\sim$}}}}
\newcommand{\bea}{\begin{eqnarray}}
\newcommand{\eea}{\end{eqnarray}}
\newcommand{\bean}{\begin{eqnarray*}}
\newcommand{\eean}{\end{eqnarray*}}
\begin{document}
\begin{titlepage}
\vspace*{-1cm}
\hfill{hep-ph/9211309}\\
\phantom{bla}
\hfill{CPTH-A206.1192}\\
\phantom{bla}
\hfill{FTUAM 92/35}\\
\phantom{bla}
\hfill{IEM-FT-63/92}\\
\phantom{bla}
\hfill{October 1992}
\vskip 0.5cm
\normalsize
\begin{center}
{\Large\bf Dynamical supersymmetry breaking with a large internal dimension
}\footnote{Work partly supported by
IN2P3-CICYT under contract PTh91-2 and by EEC contracts SC1-915053
and SC1-CT92-0792}
\end{center}
\vskip .5cm
\begin{center}
{\bf I. Antoniadis}
\\Centre de Physique Th\'eorique,
Ecole Polytechnique, 91128 Palaiseau, France\\
\vskip .3cm
{\bf C. Mu\~noz}
\\Dept. de F\'{\i}sica Te\'orica C-XI,
Univ. Aut\'onoma de Madrid, 28049 Madrid, Spain\\
\vskip .3cm
{\bf M. Quir\'os}\\
Instituto de Estructura de la Materia,
Serrano 123, 28006 Madrid, Spain
\end{center}

\begin{abstract}
\noindent
Supersymmetry breaking in string perturbation theory predicts the existence of
a new dimension at the TeV scale. The simplest realization of the minimal
supersymmetric Standard Model in the context of this mechanism has two
important consequences: (i) A natural solution to the $\mu$-problem;
(ii) The absence of quadratic divergences in the cosmological
constant, which leads to a dynamical determination of the supersymmetry
breaking and electroweak scale. We present an explicit example in which the
whole particle spectrum is given as a function of the top quark mass.
A generic prediction of this mechanism is the existence of Kaluza-Klein
excitations for gauge bosons and higgses. In particular the first
excitation of the photon could be accessible to future accelerators and
give a clear signal of the proposed mechanism.

\end{abstract}
\end{titlepage}
\nsect{Introduction}

One of the main issues in string theory is the supersymmetry
breaking, which constitutes a basic ingredient to connect the effective
field theory at the Planck scale $M_p$ to the observed low energy physics.
It
turns out that the scale of supersymmetry breaking $m_s$ does not
correspond to
an independent parameter of the compactification in four dimensions
\cite{BD}.
In every consistent perturbative solution with spontaneously broken
supersymmetry, $m_s$ is necessarily proportional to the inverse size of
some
internal dimension(s) $1/R$ \cite{ABLT,KP}. Since $m_s$ is expected to be
of the
order of the scale of weak interactions to protect the gauge hierarchy, it
follows that the decompactification scale must be in the TeV range.

{}From the field theory point of view such a large dimension would be a
theoretical disaster since gauge couplings increase, and the theory becomes
non-perturbative, very rapidly above $1/R$. More serious, the effective
theory
is non-renormalizable in that region and decoupling is generally lost. In
string
theory the latter problem is automatically solved due to the finiteness
property, while the former can be avoided in a large class of
four-dimensional
models \cite{A}. Its solution is
based on the observation that, before supersymmetry
breaking, the gauge couplings in these models do not depend on the large
dimension: the physical reason is that the Kaluza-Klein (KK) excitations
form, at each level, $N=4$ supermultiplets which give vanishing
contribution to
the $\beta$-functions. The main achievement of these string models is to
accomodate this property with that of chirality for the massless spectrum.
After
supersymmetry breaking the dependence is exponentially suppressed in the
large
radius limit \cite{A}. This phenomenon can be explicitely shown for
orbifold
compactifications. In that case the consistency of the theory after the
chiral
projection leads also to the existence of the so-called twisted states
which have
not KK-excitations.

In this class of models supersymmetry is spontaneously broken along a flat
direction, corresponding to arbitrary values of $R$, and a vanishing
vacuum energy at the tree-level. On the other hand, an important
property of this mechanism is the absence of quadratic divergences in the
cosmological constant. This allows the radiative determination of the
supersymmetry breaking and electroweak scales, as in the no-scale
supergravity
models \cite{nos}, by minimizing the full one-loop effective potential.
Both
mass-parameters are given in terms of a new scale, $Q_0$, which is
dynamically
generated through the running of the renormalization group equations (RGEs)
and
is hierarchically smaller than $M_p$.

In this paper, we work out in detail a simple realization of the minimal
supersymmetric Standard Model (MSSM) assuming a string construction which
exhibits the above properties. We find that the model is strongly
constrained
both by theoretical and phenomenological requirements. In particular,
quarks
and leptons are twisted states, while the Higgs sector plays a very special
role. In fact the second Higgs doublet, characteristic of supersymmetric
theories, is identified with the first KK-excitation of the first Higgs
doublet
carrying opposite hypercharge. This allows a natural generation of the
$\mu$-parameter in the broken supersymmetric theory, given in terms of the
compactification radius. The only non-vanishing soft breaking terms at the
Planck scale are a common gaugino mass equal to $\mu$, as well as
non-universal
Higgs masses also determined by $\mu$. After imposing all minimization
conditions one is left with one free parameter, the top Yukawa coupling
$h_t$.
We find two allowed regions: a light top region, where
$m_t\stackrel{<}{{}_\sim} 105$ GeV, and a heavy top region, where
$140\ {\rm GeV}\stackrel{<}{{}_\sim} m_t \stackrel{<}{{}_\sim} 155\ {\rm
GeV}$.
In both cases, the dynamically generated scale $Q_0$ is close to the
electroweak scale. A characteristic signature of these models is the
existence
of KK-modes of gauge bosons and higgses at low energy. It turns out
that the lightest one, an excited photon $\gamma^*$, could be accesible to
future accelerators.

To be self-contained we review, in Section 2, the main features of orbifold
compactifications that will be used, as well as the mechanism of
supersymmetry
breaking with a large internal dimension. This is the generalization of the
Scherk-Schwarz field-theoretical method \cite{SS,F} in string theory
\cite{R,KP}.
The ${\bf Z}_4$ example is worked out for pedagogical purposes. In Section
3, we
describe the dynamical determination of scales. We also compare this
approach to
the non-perturbative mechanism of supersymmetry breaking by gaugino
condensation \cite{gc}. In Section 4, we present the simplest embedding of
the
MSSM in the context of this mechanism and we explain how the $\mu$-problem
is
solved. All tree-level soft breaking parameters are obtained in terms of
one
scale $1/R$ and they are non-universal in the Higgs sector. In Section 5,
we
discuss some aspects related to the new physics of Kaluza-Klein
excitations.
In Section 6, we solve numerically the renormalization group
equations
using the above boundary conditions. We find the allowed region  consistent
with
all theoretical and experimental requirements for the only free parameter,
namely
$h_t$. The spectra of supersymmetric particles, higgses and the lowest
KK-excitation of the photon are plotted, as functions of the top-quark
mass. Finally the RGEs for couplings and soft breaking parameters of the
MSSM are collected, for completeness, in Appendix A.

\nsect{The mechanism of supersymmetry breaking}

We review here the mechanism of supersymmetry breaking with a large
internal dimension in the case of orbifold compactifications of the
heterotic
superstring.

The physical states in orbifolds can be classified in two categories
\cite{orb}:
\begin{itemize}

\item Untwisted states, which appear in the
toroidal compactification and survive after the orbifold projection.
In toroidal compactifications
every state in the higher dimensional theory gives rise
to a tower of states in four-dimensions with a mass squared shift
equal to $1/R^2$, where $R$ is the corresponding compactification
radius. These are the Kaluza-Klein excitations and their
mass comes from the components of the momentum, along the compactified
direction, which is quantized in units of $1/R$.
In string theory there are also winding states with a mass squared shift
equal to $R^2/4$, in Planck units:
\begin{equation}
M^2=M_0^2+\frac{m^2}{R^2}+ \frac{n^2R^2}{4} ,\ \ m,n=0,\pm 1,\pm 2,\ldots
 \ ,
\label{masss}
\end{equation}
where $m$ and $n$ are the momentum and winding numbers, respectively,
and $M_0$ stands for other $R$-independent contributions to the mass.
In the large radius limit we
are interested, the winding states ($n \neq 0$)
are superheavy and,
thus, irrelevant for our purposes.
In field theories toroidal compactifications lead always to
non-chiral four-dimensional spectra. In string theory, orbifolds provide
a simple way of obtaining chirality. The states from the toroidal
compactification are projected into the subset of states which are
invariant
under the discrete orbifold group.
This projection leads to a chiral massless spectrum with $N=1$
supersymmetry.

\item Twisted states, which are necessary for the internal consistency of
the
theory and correspond to string excitations around the fixed points of the
torus under the orbifold group.  Their masses do not depend on the radius
of the
torus and, therefore, they do not have KK-excitations. These states are
characteristic of string theory and do not have any analog in
field theories.
\end{itemize}

When compactifying the ten-dimensional theory to four dimensions
there appear three complex internal planes. In a general $N=1$
supersymmetric theory the Hilbert space of states can be divided into
three sectors depending on the transformation properties of the
internal planes under the action of the orbifold group.
\begin{enumerate}
\item
The $N=4$ sector, where all three planes are untwisted, which can be
seen as the truncation of a theory with $N=4$ supersymmetry.
The masses of the corresponding states depend on the deformations of all
complex planes, in particular on all internal radii. Its contribution
to threshold corrections vanishes due to the
underlying $N=4$ symmetry \cite{TV}.
\item
$N=2$ sectors, where one plane is untwisted and two planes are
twisted, which can be
seen as the truncation of a theory with $N=2$ supersymmetry.
The masses now depend only on the deformations of the untwisted
plane. As a consequence threshold corrections can depend only on
those deformations \cite{DKL}.
\item
$N=1$ sectors, where all planes are twisted.
The corresponding states and threshold corrections do not depend on any
deformation of the internal planes.
\end{enumerate}

Let us consider for simplicity only one of the three complex internal
planes with a large compactification radius $R$. To avoid troubles
with perturbation theory gauge couplings should not depend on $R$.
The simplest way is to choose a plane which is twisted in all
$N=2$ sectors.

To break spontaneously supersymmetry we use the Scherk-Schwarz
mechanism \cite{SS} that makes use of
a global, or local, continuous symmetry of the higher
dimensional theory. Upon compactification of the coordinate $X$
on a circle of radius $R$ the higher dimensional fields can be
choosen to be periodic up to a symmetry transformation:
\begin{equation}
X \rightarrow X+ 2\pi R ,\ \ \ \Phi_q \rightarrow e^{2i\pi q\omega}
\Phi_q \ ,
\label{ss}
\end{equation}
where $q$ is the charge of the field $\Phi_q$ and $\omega$ is an
arbitrary parameter of the transformation. The boundary
condition (\ref{ss}) leads to a shift in the momentum $P$ along the
$X$ direction $q \omega /R$, {\it i.e.}
\begin{equation}
P=\frac{m}{R}+\frac{q \omega}{R},\ \ \ m=0,\pm 1,\pm 2,\ldots\ ,
\label{pss}
\end{equation}
which in turn leads to a mass shift in the four-dimensional
theory. If the gravitino field has a non-zero charge
$q_{3/2}$,
supersymmetry is broken at an arbitrary scale $q_{3/2}\omega/R$.

In theories with no continuous symmetries
(left over after compactification) transforming the gravitino
field, as in the case of string theory, the above mechanism
still works for a discrete symmetry \cite{F}. In that case
the parameter $\omega$ is quantized and the scale of
supersymmetry breaking is determined only by $R$.
Note that, although supersymmetry is broken,
$R$ is undetermined at the tree level.
For the case of a ${\bf Z}_N$ symmetry,
$\omega=1$ and the charges $q$ are multiples of $1/N$.
This mechanism was extended to string theory in Ref.\cite{R,KP}.
Equations (\ref{ss}) and (\ref{pss}) remain valid for all states
with zero winding number. Consistency of the theory also
requires that the symmetry generator $q$ must have the same
transformation properties as the coordinate $X$ under the
action of the orbifold group. As a result, non-zero mass shifts
(\ref{pss}) appear only in sectors where $X$ is untwisted.

To illustrate the above ideas we present a simple example based on
the ${\bf Z}_4$ orbifold \cite{A}. The corresponding orbifold group
acting on the three complex coordinates
($X_1,X_2,X_3$), and their 2D left-moving fermionic superpartners
($\Psi_1,\Psi_2,\Psi_3$), is generated by
the element $g=(i,i,-1)$ ($g^4=1$). The only $N=2$ sector corresponds to
the
element $g^2=(-1,-1,1)$ which leaves the third internal plane
invariant. The first and second planes are therefore good candidates
to have large compactification radii.
The generator of a space-time symmetry with the desired
properties corresponds on the world-sheet to the following left-moving
complex $U(1)$ current:
\begin{equation}
J=\frac{1}{\sqrt2}(\Psi_1 \psi+ \overline{\Psi}_1
{\rm Re}\Psi_3 ) \ ,
\label{j}
\end{equation}
where $\psi^\mu$  are the left-moving 2D superpartners of
the 4D space-time coordinates $x^\mu$;
$\psi$ denotes any of the components of $\psi^\mu$.
The current (\ref{j}) does not commute with the world-sheet
supercurrent $T_F$
\begin{equation}
T_F=\psi^{\mu}\partial x_\mu +\frac{1}{2} \sum_{i=1}^3 (\Psi^i
\partial \overline{X}_i + {\rm h.c.}) \ .
\label{tf}
\end{equation}
The only allowed  discrete transformations
that commute with $T_F$ are $e^{2i\pi \oint J}$, which
leave the 2D fermions $\Psi_1,
\psi, {\rm Re}\Psi_3$ invariant.
Now we concentrate in the $N=4$ sector, which is the only one
where the coordinate $X_1$ (or $X_2$) is untwisted and therefore
we can have non-zero mass shifts. It is easy to show that the
above transformation acts on physical states as a ${\bf Z}_2$
space-time parity, namely $(-1)^{2s}$, where $s$ is the spin.
In fact, bosons come from the Neveu-Schwarz sector where 2D
fermions are anti-periodic and remain invariant under
the transformation $e^{2i\pi \oint J}$. On the other hand, fermions come
from the Ramond sector where 2D fermions
are periodic and the physical states, transforming in the spinorial
representation, change sign.
As a result, the only states which receive mass shifts from the
supersymmetry breaking mechanism are the fermions in the $N=4$
sector which survive the orbifold projection. These are the
gravitino, gauginos and all fermions in matter supermultiplets
(and their KK-excitations). Their mass is provided by
(\ref{pss}) with $\omega=1$ and ${\displaystyle q=\frac{1}{2}}$.
The general mass formula is:
\begin{equation}
M^2=M_0^2+\frac{(m+\frac{1-n}{2})^2}{R^2}
+ \frac{n^2R^2}{4},\ \ m,n=0,\pm 1,\pm 2,\ldots \ ,
\label{mass}
\end{equation}
where we also included the winding number $n$ for completeness.
As in (\ref{masss}) $M_0$ denotes a
possible contribution to the mass with a
different origin, {\it e.g.} vacuum expectation values at some scale.

All fermions in the $N=4$ sector which live in non-chiral
supermultiplets get a Majorana mass (\ref{mass}).
The mass of fermions which belong to chiral supermultiplets
looks puzzling and requires an explanation.
In fact, the supersymmetry breaking mechanism changes the
complexification of fermions within the infinite tower of
KK-excitations: the one corresponding to the
momentum number $m$ is complexified with that corresponding to
$-m-1$, as can be seen from (\ref{mass}), and becomes a Dirac
fermion. In particular the massless chiral fermion with $m=0$ is
complexified with the (anti-chiral) fermion with $m=-1$ from the
first KK-excitation and gets a Dirac mass equal to $1/2R$.
{}From the point of view of the effective field theory this
phenomenon is described by supersymmetric masses in the
superpotential,
\begin{equation}
W=\frac{1}{2R}\Phi_0\Phi_{-1} \ ,
\label{super}
\end{equation}
together with appropriate soft breaking terms for the scalar
components, such that their final masses remain unchanged:
\begin{equation}
V_{\rm sb}=-\frac{1}{4R^2}|\Phi_0|^2 + \frac{3}{4R^2}|\Phi_{-1}|^2 \ ,
\label{vsb}
\end{equation}
where $\Phi_m$ denotes the $m$-th KK-excitation, and we are using the same
symbols for superfields and their scalar components. The superpotential
(\ref{super}) leads to a fermion mass $1/2R$, while (\ref{super}) and
(\ref{vsb}) lead to scalar masses equal to zero for $\Phi_0$ and to $1/R$
for
$\Phi_{-1}$. A general consequence of
the pattern of supersymmetry breaking is that, in realistic models, quark
and
lepton multiplets should belong to twisted sectors. Otherwise they would
become
massive. Therefore these matter fields {\it do not have} light
KK-excitations.

Once supersymmetry is broken gauge couplings in general acquire
a dependence on $R$ through  radiative corrections which could
bring us back to the problem of spoiling the perturbative
expansion. However the corrections are expected to be suppressed
by the supersymmetry breaking scale, proportional to $1/R$.
In fact they are exponentially suppressed (up to constants)
\cite{A}. Moreover the radius dependence of the one-loop
cosmological constant can be calculated with the result
\cite{IT,A}:
\begin{equation}
\Lambda_{{\rm cosm}} \sim (n_B-n_F)\frac{1}{R^4}+\ldots \ ,
\label{lambda}
\end{equation}
where $n_B$ and $n_F$ are the number of massless bosons and
fermions, respectively, after the supersymmetry breaking, and
the ellipsis stands for corrections which are exponentially
suppressed in terms of $R^2$.
The absence of $1/R^2$ terms in  (\ref{lambda})
implies the vanishing of $Str {\cal M}^2$ and, hence, of quadratic
divergences in the cosmological constant.
The $1/R^4$ suppression is similar to the one obtained in $N=4$
supergravity and is a consequence of having broken supersymmetry
only in the $N=4$ sector.

Although the above discussion dealt with the ${\bf Z}_4$ orbifold,
the results apply to a more general class of models \cite{A}.
In fact, whenever we choose a radius corresponding to a plane
untwisted only in the $N=4$ sector, the pattern of supersymmetry
breaking will be the same as described above, unless the discrete
symmetry group is different from ${\bf Z}_2$. On the other hand,
there exists the possibility of choosing a plane untwisted even
in some $N=2$ sectors in models where threshold corrections do
not depend on the corresponding radius. In that case the scalar masses
from those $N=2$ sectors are shifted by the supersymmetry
breaking mechanism by the same amount as in (\ref{mass}). This
possibility is model dependent and will not be considered
in our analysis.

\nsect{Dynamical determination of scales}

The mechanism of supersymmetry breaking above described yields an
$N=1$ supersymmetric effective field theory with soft breaking terms
coming from the spontaneous breaking of supergravity.
All these terms are proportional to the inverse size
$1/R$ of the internal dimension. The soft
breaking terms can trigger radiative gauge symmetry breaking \cite{rb} at
a scale close to the supersymmetry breaking scale $m_s \sim 1/R$
when a Higgs field acquires a non-zero vacuum expectation value
(VEV) $v$. This in general happens if there is some large Yukawa
coupling $h$ in the theory, which drives the mass squared of the Higgs to
negative values.
The value of $v$ is proportional to $m_s$, which is the
only scale appearing in the renormalizable
field theory as a consequence of
the absence of quadratic divergences in the scalar masses.
Note that $R$, and so $m_s$, is arbitrary at the string tree-level
since it is given by the VEV of a
so-called modulus field $T$ which remains a flat direction at this level
after the supersymmetry breaking, while the cosmological constant vanishes.

This situation is reminiscent of the no-scale models in $N=1$ supergravity
\cite{nos}. One expects that, once supersymmetry is broken, string
radiative
corrections will lift the flatness  by generating a non trivial potential
for
the field $T$. Minimization of this potential with respect to the Higgs
field
and $T$ will fix $v$ and $R$ in terms of $M_p$ (the only scale in the
string theory). In general one would expect both scales to be of the
order of magnitude of $M_p$. However, in the absence of quadratic
divergences in the cosmological constant, there is no
$M_p^2 T^2$ term in the effective potential. So in the limit $M_p
\rightarrow \infty$ only the renormalizable part of the potential
survives and a new scale $Q_0$ is
dynamically generated through the running
of the renormalization group equations. Examples of this
mechanism have been analyzed in the past,
in the context of no-scale models \cite{nos}, and
the scale $Q_0$ was found to be
\begin{equation}
Q_0 \sim M_p e^{-\frac{4\pi}{h^2(M_p)} {\cal O}(1)} \ ,
\label{q0}
\end{equation}
where $M_p$ plays here the role of the scale fixing the boundary
conditions. Note that $Q_0$ is RGE invariant and, hence, a
new physical scale depending only on the value of the Yukawa coupling at
$M_p$. If $h$ is not too large, $Q_0$ can be hierarchically smaller
than $M_p$. Both $v$ and $m_s$ are then proportional and of the order
of magnitude of $Q_0$; this could explain dynamically
the hierarchy between the weak and Planck scales without
any fine-tuning.

The class of string models we have introduced in the previous section
provides good candidates for a dynamical generation of the $Q_0$
scale, since the condition for the vanishing of $Str {\cal M}^2$ is
fulfilled as can be seen from (\ref{lambda}). We stress that the
determination of $Q_0$ has a perturbative nature since it relies upon
the one-loop effective potential which is calculable in string
theory. This is in contrast with non-perturbative mechanisms of
supersymmetry breaking, {\it e.g.} gaugino condensation in the hidden
sector \cite{gc}. In the latter, there is a non-perturbatively
generated condensation scale $\Lambda$ which triggers a supersymmetry
breaking scale in the observable sector through gravitational
interactions $m_s \sim \Lambda^3/M_p^2$. This method, since it
relies on non-perturbative effects, suffers from a partial
lack of calculability at the present state of the art.
The situation becomes worse in the context of string theory, where
there is no known framework of studying non-perturbative phenomena.
Another important difference between the two mechanisms is that the
value of $Q_0$ is controled by parameters that can be measured at low
energy, while $\Lambda$ is controled by the hidden sector of the
theory, which cannot be directly measured by present experiments.

In 4D strings all coupling constants depend on the VEV's of
scalar fields, the moduli, which determine the shape and size of
the internal space. Their values should be determined
dynamically by the minimization of the effective potential. As long
as supersymmetry remains unbroken they are undetermined since the
moduli are exact flat directions (in perturbation theory). Once
supersymmetry is broken the degeneracy is lifted by the generated
potential. An important property of string theory,
in the absence of supersymmetry breaking, is the duality
symmetry which relates small and large radii by exchanging momentum
and winding modes. In the simplest case of one radius,
as can be seen from (\ref{masss}), the spectrum is
invariant under ${\displaystyle R \rightarrow \frac{2}{R}}$ and
$n \leftrightarrow m$. This symmetry is expected to remain at the level of
the effective field theory of massless modes ($m=n=0$) even in the
presence of gaugino condensation. This implies that self-dual points,
{\it e.g.} $R=\sqrt{2}$, are always extrema of the potential, and
possibly minima unless duality is spontaneously broken. This
result does not necessarily requires that gaugino condensation breaks
supersymmetry. If supersymmetry is broken perturbatively by a large
radius a modified duality transformation still remains as an invariance
of the string theory. As can be seen from (\ref{mass}) the shifted
spectrum is invariant under
${\displaystyle R \rightarrow \frac{1}{R}}$ and
$n \leftrightarrow (2m+1-n)$. In this case the effective field theory
of originally massless states is not any longer
duality invariant. For instance, under this transformation the
gauginos ($m=n=0$) are related to superheavy winding states
($m=0,n=1$) which are not included in the effective theory.
As a result the self-dual point $R=1$ is not in general an extremum
of the potential and, as we discussed before, $R$ can be fixed at a
very large value related to a new dynamical scale $Q_0$.
Radiative corrections in the supersymmetry broken theory are expected
to generate also a potential for all other moduli unrelated to $R$,
lifting their degeneracy. For them the effective theory remains
invariant under the corresponding duality transformations and
self-dual points will be extrema of the potential.

In addition to the moduli there is a universal scalar field, the
dilaton, which plays a very particular role because its VEV
determines the tree-level 4D string coupling constant.
It is clear that the VEV of this field cannot be fixed at any given
order of string perturbation theory, unless some resummation could be
performed. The situation is not improved even
in the presence of one-gaugino condensation, where a runaway
potential is obtained for the dilaton. This problem can
be fixed if two different condensates are present \cite{2gc}.
A possible scenario would be fixing the dilaton and moduli which
appear in the gauge couplings by such non-perturbative effects
without breaking supersymmetry, while supersymmetry is broken by some
of the remaining radii using the perturbative approach described
above.

\nsect{The minimal embedding of the Standard Model and a solution to the
$\mu$-problem}

In this section we present the effective theory corresponding to the
simplest embedding of the minimal supersymmetric
Standard Model in the general context described above.

As we already pointed out in Section 2, all gauginos get a common Majorana
mass
given by (\ref{mass}),
\begin{equation}
M_{1/2} = \frac{1}{2R},
\label{mg}
\end{equation}
since they belong to non-chiral supermultiplets from the untwisted sector.
On
the other hand quarks and leptons should be identified with twisted states,
having no light KK-excitations. Thus, all their corresponding soft breaking
scalar masses vanish,
\begin{equation}
m_{\tilde q} = m_{\tilde \ell} = 0.
\label{mq}
\end{equation}

In the MSSM one needs a Higgs sector
with two doublets, $H_1$ coupled to down quarks and leptons, and $H_2$
coupled
to up quarks. These couplings exhibit a global Peccei-Quinn
symmetry, which is spontaneously broken along with the electroweak
symmetry,
giving rise to a massless axion. To avoid this problem one has to introduce
a
supersymmetric mass-term in the superpotential $\mu H_1 H_2$, where $\mu$
must
be of the order of the electroweak scale. The introduction of such a
parameter looks completely unnatural, since it brings back the hierarchy
problem at tree-level even for unbroken supersymmetry. Its origin
requires an explanation at the more fundamental level. In fact, if the
higgses
belong to the untwisted sector, the mechanism of supersymmetry breaking we
described above provides a natural solution to the $\mu$ problem. Following
Section 2, one can identify one of the higgses with the first KK-excitation
of
the other. Then the Higgs superpotential and soft breaking mass terms are
given
by (\ref{super}) and (\ref{vsb}). As a consequence, one obtains
\begin{equation}
\mu = \frac{1}{2R}.
\label{mu}
\end{equation}
Depending on which Higgs field is assigned to the first KK-excitation, one
obtains two distinct cases of Higgs masses:
\begin{equation}
\begin{array}{rlllll}
{\rm case} \ 1:\ \ \ \ \ &H_1 = \Phi_{-1} \ &H_2 = \Phi_0
\ &\Rightarrow \ &m_1 = \frac{1}{R} \ &m_2 = 0 \\
{\rm case} \ 2:\ \ \ \ \ &H_2 = \Phi_{-1} \ &H_1 = \Phi_0
\ &\Rightarrow \ &m_1 = 0 \ &m_2 = \frac{1}{R}.
\end{array}
\label{cases}
\end{equation}
Note that in (\ref{vsb}) there is no soft breaking mixing between the two
higgses, $B\mu \Phi_0 \Phi_{-1}$, implying that the soft breaking parameter
$B$
vanishes,
\begin{equation}
B = 0.
\label{B}
\end{equation}

In both cases (\ref{cases}) we use only one Higgs doublet that was massless
before supersymmetry breaking. This raises the problem of anomaly
cancellation in the supersymmetric theory. Since all massless fermions from
the
untwisted sector become massive in the broken theory, while those from the
twisted sectors remain massless, the anomaly cancellation condition after
supersymmetry breaking only involves the twisted states. Therefore, there
are
two possible scenarios consistent at the string level:
\begin{enumerate}
\item
The supersymmetric theory is anomaly free, implying that
$U(1)_Y$-hypercharge
anomalies must cancel separately in untwisted and twisted sectors. In this
case
there should exist extra untwisted matter representations. The simplest
possibility consists in having an additional Higgs
doublet with no coupling to quarks and leptons.
\item
The supersymmetric theory has only one massless Higgs doublet and the
$U(1)_Y$ is anomalous. As all $SU(2)$ and gravitational mixed
anomalies are proportional to Tr$Y$, this anomaly can be cancelled by a
Green-Schwarz counterterm \cite{u1a}. Although the supersymmetric theory
appears
to be unrealistic since the hypercharge would be broken at the Planck
scale,
in the broken theory the $U(1)_Y$ becomes anomaly free and one is left with
the
minimal supersymmetric standard model. This unconventional possibility
could
be interesting for model building.
\end{enumerate}

It remains to study the $A$ soft breaking parameters, which correspond to
the
trilinear couplings in the scalar potential. They are given by the on-shell
tree-level amplitude involving three complex scalars whose fermionic
partners
have the same chirality. This amplitude vanishes as in the supersymmetric
theory, since the scalar vertex operators remain unchanged after
supersymmetry
breaking. Therefore
\begin{equation}
A = 0.
\label{A}
\end{equation}

To summarize, the simplest embedding of the MSSM we have just described
leads to
the tree-level values of masses and soft-breaking parameters given in
(\ref{mg}-\ref{A}). They are {\it non-universal} in the Higgs sector
(\ref{cases}),  and they are given in terms of one mass scale, namely the
inverse size of the extra dimension $1/R$. The latter corresponds to the
VEV of
the modulus field $T$ which is a flat direction of the tree-level
potential. As
we explained in Section 3, the value of $1/R$ along with the VEVs of the
two
Higgs doublets should be dynamically determined by minimizing the one-loop
effective potential with respect to these fields. In the leading
logarithmic
approximation, this is equivalent to considering the renormalization group
equations of the MSSM with boundary conditions defined by
(\ref{mg}-\ref{A}). The resulting VEVs will then be proportional to the
RGE-invariant scale $Q_0$, defined in (\ref{q0}) as the scale where the
determinant of the Higgs mass-squared matrix becomes negative and triggers
electroweak symmetry breaking:
\begin{equation}
\det {\cal M}_H^2(Q_0) = 0.
\label{q0def}
\end{equation}
As a consequence of the vanishing of all quadratic divergences, $Q_0$ is
independent of $M_p$ and can be hierarchically smaller depending on the
largest Yukawa coupling, namely that of the top quark $h_t$. Note that
$Q_0$
is also independent of the supersymmetry breaking scale $1/R$. The
minimization of the effective potential with respect to the Higgs fields
$H_1$ and $H_2$ fixes their VEVs in terms of $1/R$, while the minimization
with respect to $T$ fixes $1/R$ in terms of $Q_0$. As a result, all
low-energy masses are fixed in terms of $h_t$. Any
particular string model which provides a definite prediction for $h_t$ will
lead to a precise prediction for the low-energy mass spectrum of the
theory.
Here, since we are not dealing with a particular model, we will consider
$h_t$ as
a free parameter, and we will replace the minimization with respect to $T$
by
the phenomenological condition on the value of the $Z$-mass. In Section 6
the mass-spectrum will be derived as a function of the top-quark mass
$m_t$.
As a consistency check of the procedure, one should verify that the
obtained
value for $Q_0$ is of the order of the scale of weak interactions.

\nsect{The physics of Kaluza-Klein excitations}

An unavoidable consequence of the
proposed mechanism of supersymmetry breaking with a large internal
dimension is
the existence of KK-excitations at low energy. The lightest of such states
is
the first excitation of the photon, $\gamma^*$, with a mass
\begin{equation}
M_{\gamma^*} = \frac{1}{R}
\label{mgamma}
\end{equation}
accessible to future accelerators, and a very clear signal in the
$\ell^+\ell^-$
channel. Here, we will discuss a few theoretical aspects of their
interactions,
as well as possible effects and constraints at present energies that could
lead
to bounds or signatures of them.

As we have seen in the previous sections, only the untwisted states have
KK-excitations, namely all gauge boson and Higgs supermultiplets.
Although, by construction, these KK-modes do not affect the running of
coupling
constants, they could lead to observable effects through non-renormalizable
interactions. In a supersymmetric theory, such higher dimensional
interactions
may come from $F$ or $D$-terms. Since quarks and leptons are in the
twisted
sector, the interactions corresponding to $F$-terms involve more than three
twisted fields and, thus, are exponentially suppressed in the large radius
limit \cite{CV}. This does not apply to $D$-terms which can lead to
dimension six
effective operators corresponding, for instance, to the exchange of massive
vector bosons.

In the $N=4$ supersymmetric theory obtained from a toroidal
compactification, all KK-excitations have the same quantum numbers as those
of
the lowest lying massless states. At the massless level, some quantum
numbers
are projected out because they are not invariant under the orbifold group.
However, they all appear at the massive level because one can always
construct
linear combinations of KK-modes which are invariant under the orbifold
group,
and carry these quantum numbers. A simple example of this phenomenon is the
hypercharge in the Higgs sector discussed in Section 4: the massive
excitations
of one Higgs doublet contain doublets with the opposite hypercharge. As a
result,
in a general $N=1$ theory, KK-excitations may have additional quantum
numbers
compared to those of the lowest lying massless states. This might give rise
to
unwanted, phenomenologically suppressed processes, as fast proton decay.
However, they can be easily avoided if the model has no massless color
triplets at
the $N=4$ level, {\it i.e.} before the orbifold projection is applied.

Now, we concentrate on model independent dimension six operators which are
produced by the exchange of KK-modes with the same quantum numbers as the
lowest
lying states ({\it e.g.} $\gamma^*$, $Z^*$, \dots). Either they modify the
effective parameters of the Standard Model ({\it e.g.} $Z^*$, $W^*$), or
they produce new effective interactions ({\it e.g.}
$\gamma^*$, excited gluon). In
the former case, the mass of the excited modes is bounded by the precision
measurements of weak interactions. In the latter, the mass is constrained
by
bounds on new physics, obtained from those on compositeness. The general
form
of such operators can be written as:
\begin{equation}
\frac{e^2}{2M^2}({\overline\psi}\gamma^\mu \psi)^2 \ ,
\label{oper}
\end{equation}
where $M$ is the mass of the intermediate state and $e$ its coupling to the
massless fermions $\psi$ at low energy. The tree-level coupling of a
massive
KK-mode, corresponding to the momentum number $m$ (see (\ref{masss})), to
two
massless twisted states is proportional to $\delta^{-m^2/2R^2}$, where
$\delta$
is a number depending on the orbifold twist \cite{DFMS}. In the large
radius
limit, this coefficient is equal to one, and the coupling is independent of
$m$
and equal to the corresponding coupling of the lowest lying massless state,
$m=0$. To determine $e$ in (\ref{oper}) at low energy, we need also to
study the
infrared running of the coupling of the massive KK-mode from the Planck
scale to
its mass. In this region, this mode can be treated as massless and its
vertex
coincides with that of the corresponding massless state. Therefore, the one
loop string diagrams will have the same infrared divergence and both
couplings
will be renormalized in the same way.

To extract a first estimate on present bounds on the size of the internal
dimension in the above context, we will use the experimental limits on the
strength of four-fermion interactions \cite{PDG}. The stronger bound comes
from $ee\mu\mu$ which gives:
\begin{equation}
R^{-1} \simgt \frac{\sqrt 6}{\pi} \frac{1}{2.2} \alpha_{\rm em}^{1/2}
\Lambda_{ee\mu\mu} \ ,
\label{bound}
\end{equation}
where $\Lambda_{ee\mu\mu}$ is the compositeness scale. The first factor in
the
r.h.s. of (\ref{bound}) comes from the sum over the infinite tower of
massive
photons, while the second factor comes from the contribution of the excited
$Z$'s \cite{KS}. Using the present value of $\Lambda_{ee\mu\mu} \sim 4.4$
TeV \cite{PDG},
one obtains $R^{-1} > 140$ GeV. This means that $\gamma^*$ could still be
detected even at LEP-200!

A number of interesting theoretical questions is related to the physics of
KK-modes at high energies. As the energy increases above the
decompactification scale, one produces more states which behave
effectively as massless vector particles. At the field theory level, a
higher
gauge symmetry is required to describe consistently their interactions, and
a
Higgs mechanism is needed to explain the origin of their masses. This
symmetry
should be a small part of the full string symmetry, which is infinite
dimensional in the limit of Planckian energies \cite{GP}. In this sense,
the
detection of these particles and the understanding of their interactions
could
be considered as a window to uncover the complicated underlying string
structure. A related issue is the possibility of observation of the extra
dimension in a reaction where some outgoing particles could ``propagate" in
the
decompactified space. This naive field-theoretical picture is incorrect
because there is no conserved quantum number associated to this dimension,
even
approximately. In fact, as we discussed, all KK-modes are unstable and
decay to
massless states from the twisted sector. This is a stringy phenomenon
arising
from the chiral character of the theory and implying that space-time
behaves,
below the Planck scale, always as four-dimensional from the observational
point
of view.

\nsect{The low energy particle spectrum}

Let us analyze now the radiative breaking of the
$SU(2)\times U(1)_Y$ symmetry \cite{rb}
triggered by the mechanism of spontaneous supersymmetry
breaking described above. A main difference with respect to the
usual
low-energy supergravity models is the presence of the infinite tower
of massive excitations. However, as we already pointed out, these modes do
not
affect the evolution of gauge couplings because they form $N=4$ multiplets.
For
the same reason their contribution to wave function renormalization of
Higgs and
matter multiplets also vanishes. On the other hand, since heavier Higgs
excitations have the same coupling to matter as the lightest one, they
cannot get any VEV and, thus, they do not affect the minimization of the
effective potential. Therefore, all higher KK-modes can be neglected in the
evolution of RGEs. Our analysis will then be similar to the usual one after
taking into account the (non-universal) boundary conditions
(\ref{mg}-\ref{A})
we found in Section 4.

For completeness, we give in Appendix A the RGEs for all the parameters of
the
MSSM \cite{jap,S}. Note that in these equations there is a contribution to
the
scalar masses coming from the $U(1)_Y$ $D$-term parametrized in the general
case
by:
\begin{equation}
S = \sum_r d(r) Y_r m_r^2,
\label{S}
\end{equation}
where $d(r)$ is the dimension of the representation $r$. The evolution
of $S$ is given by \cite{S}:
\begin{equation}
\frac{dS}{dt}=-\frac{\alpha_1}{4\pi} (\sum_r d(r) Y_r^2) \ S,
\label{Seq}
\end{equation}
implying that if $S$ vanishes at some scale, it vanishes everywhere.
For the case of universal boundary conditions for the scalar masses, $S=0$.
In our case, $S$ seems to be non-vanishing at the string scale due
to the non-universality of boundary conditions in the Higgs sector
(\ref{cases}). However, this is an artifact of our truncation:
taking into account the contribution to $S$ of the tower of KK-excitations,
one
finds at each level an equal number of massive scalars with opposite
hypercharges except at the lowest level which is massless. This implies
that $S$ is in fact vanishing at the string scale.

Neglecting all Yukawa couplings except the one of the top, we have in
principle two free parameters at the string scale:
\begin{equation}
h_t,\;\;\mu
\label{alfexp}
\end{equation}
In order to obtain the $SU(2)\times U(1)_Y$ breaking at the right
scale we must impose
\begin{equation}
M_{Z}^2=\frac{1}{2}(g_2^2+g_1^2)(v_1^2+v_2^2) \ ,
\label{MZ}
\end{equation}
where $g_2$ and $g_1$ are the $SU(2)$ and $U(1)_Y$ gauge couplings,
respectively, and
$v_{1,2} \equiv \newline <H_{1,2}>$. In this way, we are left with only one
free parameter and the supersymmetric spectrum will be strongly
constrained.

The relevant Higgs scalar potential along the neutral direction is
\begin{equation}
V(H_1,H_2)=\frac{1}{8}(g_2^2+g_1^2)\left(|H_1|^2-|H_2|^2\right)^2
+ m_1^2|H_1|^2 + m_2^2|H_2|^2 - m_3^2(H_1H_2+
{\rm h.c.})
\label{Vhiggs}
\end{equation}
where $m_1$, $m_2$, $m_3$ are related to the soft breaking parameters
via expressions coming from integrating the corresponding RGEs.
After integration, all renormalized soft parameters are
proportional to $\mu$. In particular,
\begin{equation}
\begin{array}{ll}
m_1^2=(l(t) + g(t) + a)\mu^2 &\equiv C_1\mu^2 \\
m_2^2=(l(t) + e(t) + b)\mu^2 &\equiv C_2\mu^2 \\
m_3^2=r(t)\mu^2 &\equiv C_3\mu^2 ,
\end{array}
\label{mus1}
\end{equation}
where the functions $l(t)$, $g(t)$, $e(t)$ and $r(t)$ can be found in
\cite{carlos}. They depend on the parameter $t=2\log(M_{SU}/M_Z)$, where
$M_{SU}$ is the string unification scale, as well as on $h_t$ and the gauge
coupling constants $\alpha_i$. The constants $a$ and $b$ appearing in
(\ref{mus1}) are $a=3(-1)$, $b=-1(3)$ for the case 1 (case 2).
The minimization conditions of the scalar
potential (\ref{Vhiggs}) together with the constraint (\ref{MZ}) give:
\begin{eqnarray}
\omega^2 &={\displaystyle
\frac{m_1^2+\frac{1}{2}M_Z^2}{m_2^2+\frac{1}{2}M_Z^2}}
\label{mini1} \ , \\
\frac{\omega}{\omega^2+1} &={\displaystyle \frac{m_3^2}{m_1^2+m_2^2}} \ ,
\label{natur2}
\end{eqnarray}
with $\omega \equiv{\displaystyle \frac{v_1}{v_2}}$. Combining
(\ref{mini1}) and (\ref{natur2}), and using (\ref{mus1}), we determine the
value
of $\mu$:
\begin{equation}
\mu=\left\{ \frac{M_Z^2\left[ (C_1+C_2)^2-4C_3^2\pm
(C_1-C_2)\sqrt{(C_1+C_2)^2-4C_3^2} \right]}{4(C_1+C_2)(C_3^2-C_1C_2)}
 \right\}^{\frac{1}{2}}.
\label{Mchar}
\end{equation}
It turns out that for the boundary conditions
we use, the solution corresponding to the lower sign in (\ref{Mchar}) is
unrealistic because it gives rise to imaginary values for $\mu$. We are
left
with only one free parameter $h_t$ and the whole supersymmetric mass
spectrum
can be obtained as a function of the top mass $m_t$.

We have solved numerically the RGEs of Appendix A in the approximation
$h_b$,
$h_{\tau}\ll h_t$ and computed the
supersymmetric particle masses by means of the usual formulae \cite{rb}.
All the masses are evaluated at the $M_Z$ scale and the  running is
done from $M_{SU}\simeq 0.5\times g \times 10^{18}$ GeV \cite{K}, where
$g\simeq 1/\sqrt{2}$ is the corresponding value of the string coupling
constant\footnote{We have checked that the inclusion
of $h_b$, $h_{\tau}$ in the calculation does not
modify essentially our results. The same occurs if we
do the running from $M_{GUT}$ instead of $M_{SU}$.}.
The current experimental values that we use are:
\begin{equation}
M_Z=91.175,\;\;\alpha_3^{-1}(M_Z)=8,\;\;\alpha_{\rm em}^{-1}(M_Z)=127.9,
\;\;\sin^2\theta_W(M_Z)=0.2303 \ .
\label{alfexp}
\end{equation}
$\alpha_i$ at $M_{SU}$ should be obtained by inserting $\alpha_i(M_Z)$
in the corresponding RGEs.

The numerical results are summarized in figs. 1 and 2, corresponding to the
boundary conditions of cases 1 and 2 in (\ref{cases}), respectively, where
the
whole spectrum is plotted as a function of $m_t$. It is worth noticing that
although we have only one free parameter, there are still solutions
consistent
with the experimental bounds. This is certainly non-trivial and the allowed
solutions strongly constrain the range of the top quark mass. In fact, case
1
leads to a {\it light} top in the range
\begin{equation}
90\ {\rm GeV} < m_t \stackrel{<}{{}_\sim} 105\ {\rm GeV},
\label{Seq}
\end{equation}
which corresponds to a range of the top Yukawa coupling $h_t$ at the string
scale between 0.14 and 0.18. The lower bound comes from the present
experimental
limit, whereas the upper bound is due to the sneutrino mass which becomes
negative for $m_t>105$ GeV (see fig.1b). The values of $\omega$ can be read
off
from the top of the plots and they vary in the range between 10.6 and 12.9.
Fig.1a shows squark ($\tilde{q}$) and
gluino ($\tilde{g}$) masses. In our approximation u and d-type
squarks are  degenerate for the first two generations (separately for each
chirality and charge), and the right sbottom is also degenerate with the
first
two right d-squarks. Left and right stops have a non-trivial mixing in
their mass
matrix which gives rise to the eigenstates  $\tilde{t}_\ell$ and
$\tilde{t}_h$.
Fig.1b shows Higgs, slepton ($\tilde{\ell}$) and excited photon
($\gamma^*$)
masses. In the usual notation
$H^{\pm}$ is the charged Higgs, $A$ is the pseudoscalar, and $H$, $h$ are
the neutral scalars. Again in our approximation, all sleptons are
degenerate
(separately for each chirality and charge). Fig.1c shows the
neutralino ($\chi^o$) and chargino ($\chi^{\pm}$) masses. Note that the
masses of sneutrinos, right selectrons, lightest chargino and the two
lightest
neutralinos are close to their present experimental bounds \cite{PDG}. The
lightest supersymmetric particle (LSP) is a neutralino for $m_t<102$ GeV
and
the sneutrino for larger values. Finally, the values of the dynamically
generated scale $Q_0$ defined in (\ref{q0def}) are in the range 100-700
GeV, justifying the reliability of the RGE treatment.

The boundary conditions of case 2 in (\ref{cases}) lead to a {\it heavy}
top
in the range:
\begin{equation}
140\ {\rm GeV} \stackrel{<}{{}_\sim} m_t \stackrel{<}{{}_\sim} 155\ {\rm
GeV},
\label{Seq}
\end{equation}
which corresponds to a range of the top Yukawa coupling $h_t$ at the string
scale between 0.4 and 0.38. The lower bound in (\ref{Seq}) comes from the
present experimental limits on supersymmetric Higgs detection: for $m_t <
140$
GeV the lightest neutral Higgs $h$ becomes too light as shown in fig.2b.
Note that radiative corrections to its mass are important in this case, and
amount to adding around 20 GeV to the tree-level values
\cite{radcor}. At the upper bound of
$m_t$ the value of $\mu$ in  (\ref{Mchar}) goes to infinity, and beyond it
becomes imaginary. However, in this region one should take into account the
decoupling of heavy particles in the RGEs to obtain a
more reliable result. The
values of $\omega$ are now in the range between 1.7 and 4, while $Q_0$
varies
between 140 and 100 GeV. In this case the LSP is a neutralino in the whole
range
of allowed $m_t$.

In conclusion, we have presented the phenomenology of the minimal
supersymmetric Standard Model in the context of a class of 4D strings with
spontaneously broken supersymmetry by a large internal dimension. The
scales
of supersymmetry breaking and weak interactions are proportional to a
unique
dynamically determined scale hierarchically smaller than the Planck mass,
and
the whole spectrum was given as a function of one free parameter, the top
quark
mass. The boundary conditions at the string unification scale were
determined by
the supersymmetry breaking mechanism applied in a particular class of
compactifications which include the ${\bf Z}_4$ orbifold. The
classification of
other possible boundary conditions, arising from different exact discrete
symmetries used to break supersymmetry, is an interesting question. On the
other hand, the explicit construction of a ``realistic" string model
exhibiting the required properties remains an open problem.

\begin{appendix}
\nsect{Appendix}

We collect here the renormalization group equations for the couplings and
soft
breaking parameters of the minimal supersymmetric Standard Model. We
neglect all
Yukawa couplings except those of the third generation.
\begin{itemize}
\item
Gauge couplings
\begin{eqnarray}
\frac{dg_i^2}{dt}=-\frac{b_i}{(4\pi)^2}g_i^4
\label{gauge}
\end{eqnarray}
\item
Gaugino masses
\begin{eqnarray}
\frac{dM_i}{dt}=-\frac{b_i}{(4\pi)^2}g_i^2 M_i^2
\label{gauginos}
\end{eqnarray}
where $t \equiv ln(M_{GUT}^2/Q^2)$ and $b_3=-3$, $b_2=1$ and $b_1=11$.
\item
Yukawa couplings of the third generation
\begin{eqnarray}
\frac{dY_t}{dt}=Y_t(\frac{16}{3}\tilde\alpha_3+3\tilde\alpha_2+
\frac{13}{9}\tilde\alpha_1-
6Y_t-Y_b)
\label{Yuk1}
\end{eqnarray}

\begin{eqnarray}
\frac{dY_b}{dt}=Y_b(\frac{16}{3}\tilde\alpha_3+3\tilde\alpha_2+
\frac{7}{9}\tilde\alpha_1-
Y_t-6Y_b-Y_{\tau})
\label{Yuk2}
\end{eqnarray}

\begin{eqnarray}
\frac{dY_{\tau}}{dt}=Y_{\tau}(3\tilde\alpha_2+3\tilde\alpha_1-3Y_b-4Y_{\tau}
)
\label{Yuk3}
\end{eqnarray}
where one defines
\begin{eqnarray}
Y \equiv \frac{h^2}{(4\pi)^2};\;\;\;\tilde\alpha_i \equiv
\frac{\alpha_i}{4\pi}
\label{def}
\end{eqnarray}
and the gauge coupling constants at $M_{GUT}$ verify
\begin{equation}
\alpha_3(0)=\alpha_2(0)=\frac{5}{3}\alpha_1(0)=\alpha_{GUT}
\end{equation}
\item
Trilinear soft terms corresponding to the third generation Yukawa couplings
\begin{eqnarray}
\frac{dA_t}{dt}=(\frac{16}{3}\tilde\alpha_3M_3+3\tilde\alpha_2M_2+
\frac{13}{9}\tilde\alpha_1M_1)-6Y_tA_t-Y_bA_b
\label{A1}
\end{eqnarray}

\begin{eqnarray}
\frac{dA_b}{dt}=(\frac{16}{3}\tilde\alpha_3M_3+3\tilde\alpha_2M_2+
\frac{7}{9}\tilde\alpha_1M_1)-Y_tA_t-6Y_bA_b-Y_{\tau}A_{\tau}
\label{A2}
\end{eqnarray}

\begin{eqnarray}
\frac{dA_{\tau}}{dt}=(3\tilde\alpha_2M_2+3\tilde\alpha_1M_1)-
3Y_bA_b-4Y_{\tau}A_{\tau}
\label{A3}
\end{eqnarray}
\item
Supersymmetric mass of the higgsinos
\begin{eqnarray}
\frac{d\mu^2}{dt}=(3\tilde\alpha_2+\tilde\alpha_1-3Y_t-3Y_b-Y_{\tau})\mu^2
\label{mu}
\end{eqnarray}
\item
Scalar masses
\begin{equation}
\begin{array}{rl}
{\displaystyle
\frac{dm_Q^2}{dt} }&{\displaystyle
=(\frac{16}{3}\tilde\alpha_3M_3^2+3\tilde\alpha_2M_2^2+
\frac{1}{9}\tilde\alpha_1M_1^2)
- Y_t(m_Q^2+m_u^2+m_2^2+A_t^2-\mu^2)}\\
&-{\displaystyle
Y_b(m_Q^2+m_d^2+m_1^2+A_b^2-\mu^2)-\frac{1}{6}\tilde\alpha_1S }
\label{mas1}
\end{array}
\end{equation}

\begin{equation}
\frac{dm_u^2}{dt}
=(\frac{16}{3}\tilde\alpha_3M_3^2+\frac{16}{9}
\tilde\alpha_1M_1^2)-
2Y_t(m_Q^2+m_u^2+m_2^2+A_t^2-\mu^2)
+\frac{2}{3}\tilde\alpha_1S
\label{mas2}
\end{equation}

\begin{equation}
\frac{dm_d^2}{dt}=(\frac{16}{3}\tilde\alpha_3M_3^2+\frac{4}{9}
\tilde\alpha_1M_1^2)-
2Y_b(m_Q^2+m_d^2+m_1^2+A_b^2-\mu^2)-\frac{1}{3}\tilde\alpha_1S
\label{mas3}
\end{equation}

\begin{eqnarray}
\frac{dm_L^2}{dt}=(3\tilde\alpha_2M_2^2+\tilde\alpha_1M_1^2)-
Y_{\tau}(m_L^2+m_e^2+m_1^2+A_{\tau}^2-\mu^2)+\frac{1}{2}\tilde\alpha_1S
\label{mas4}
\end{eqnarray}

\begin{eqnarray}
\frac{dm_e^2}{dt}=(4\tilde\alpha_1M_1^2)-
2Y_{\tau}(m_L^2+m_e^2+m_1^2+A_{\tau}^2-\mu^2)-\tilde\alpha_1S
\label{mas5}
\end{eqnarray}
For the scalars of the first two generations the same equations apply with
$Y_t=Y_b=Y_{\tau}=0$.
\item
Higgs doublets mass parameters
\begin{equation}
\begin{array}{rl}
{\displaystyle
\frac{dm_1^2}{dt}}&{\displaystyle
=(3\tilde\alpha_2M_2^2+\tilde\alpha_1M_1^2)+(3\tilde\alpha_2+
\tilde\alpha_1)\mu^2
-3Y_t\mu^2-3Y_b(m_Q^2+m_d^2+m_1^2+A_b^2) }\\
&-{\displaystyle
Y_{\tau}(m_L^2+m_e^2+m_1^2+A_{\tau}^2)+\frac{1}{2}\tilde\alpha_1S }
\label{m1}
\end{array}
\end{equation}

\begin{equation}
\begin{array}{rl}
{\displaystyle
\frac{dm_2^2}{dt}}&{\displaystyle
=(3\tilde\alpha_2M_2^2+\tilde\alpha_1M_1^2)+(3\tilde\alpha_
2+
\tilde\alpha_1)\mu^2 }\\
&{\displaystyle
-(3Y_b+Y_{\tau})\mu^2-3Y_t(m_Q^2+m_u^2+m_2^2+A_t^2)-
\frac{1}{2}\tilde\alpha_1S }
\label{m2}
\end{array}
\end{equation}

\begin{equation}
\begin{array}{rl}
{\displaystyle
\frac{dm_3^2}{dt}}&{\displaystyle
=(\frac{3}{2}\tilde\alpha_2+\frac{1}{2}
\tilde\alpha_1-\frac{3}{2}Y_t
-\frac{3}{2}Y_b-\frac{1}{2}Y_{\tau})m_3^2 }\\
&{\displaystyle
-(3\tilde\alpha_2M_2+\tilde\alpha_1M_1)\mu
+(3Y_tA_t+3Y_bA_b+Y_{\tau}A_{\tau})\mu }
\label{m3}
\end{array}
\end{equation}
where the value of S is given by
\begin{eqnarray}
S=m_2^2-m_1^2+\sum_{\rm generations}(m_Q^2+m_d^2-2m_u^2-m_L^2+m_e^2)
\label{mas5}
\end{eqnarray}
\end{itemize}
\end{appendix}

\nsect*{Figure captions}

\begin{description}
\item[Fig.1:]
Masses of the different
particles versus $m_t$ for the case 1: a) Squarks and gluinos;
b) Higgses, sleptons and excited photon; c) Neutralinos and charginos.
\item[Fig.2:]
The same as in fig.1 but for the case 2.
\end{description}

\end{document}